\documentclass[dvips]{article}
\usepackage{supertabular,lscape,epsfig}
\usepackage{amssymb}
\usepackage{amsmath}

\DeclareSymbolFont{ppa}{OT1}{ppl}{m}{it}
\DeclareMathSymbol{\vv}{\mathalpha}{ppa}{'166}

\begin{document}

\newcommand{\dd}{\,{\rm d}}
\newcommand{\ie}{{\it i.e.},\,}
\newcommand{\etal}{{\it et al.\ }}
\newcommand{\eg}{{\it e.g.},\,}
\newcommand{\cf}{{\it cf.\ }}
\newcommand{\vs}{{\it vs.\ }}
\newcommand{\zdot}{\makebox[0pt][l]{.}}
\newcommand{\up}[1]{\ifmmode^{\rm #1}\else$^{\rm #1}$\fi}
\newcommand{\dn}[1]{\ifmmode_{\rm #1}\else$_{\rm #1}$\fi}
\newcommand{\upd}{\up{d}}
\newcommand{\uph}{\up{h}}
\newcommand{\upm}{\up{m}}
\newcommand{\ups}{\up{s}}
\newcommand{\arcd}{\ifmmode^{\circ}\else$^{\circ}$\fi}
\newcommand{\arcm}{\ifmmode{'}\else$'$\fi}
\newcommand{\arcs}{\ifmmode{''}\else$''$\fi}
\newcommand{\MS}{{\rm M}\ifmmode_{\odot}\else$_{\odot}$\fi}
\newcommand{\RS}{{\rm R}\ifmmode_{\odot}\else$_{\odot}$\fi}
\newcommand{\LS}{{\rm L}\ifmmode_{\odot}\else$_{\odot}$\fi}

\newcommand{\Abstract}[2]{{\footnotesize\begin{center}ABSTRACT\end{center}

\vspace{1mm}\par#1\par   
\noindent
{~}{\it #2}}}

\newcommand{\TabCap}[2]{\begin{center}\parbox[t]{#1}{\begin{center}
  \small {\spaceskip 2pt plus 1pt minus 1pt T a b l e}
  \refstepcounter{table}\thetable \\[2mm]
  \footnotesize #2 \end{center}}\end{center}}

\newcommand{\TableSep}[2]{\begin{table}[p]\vspace{#1}
\TabCap{#2}\end{table}}

\newcommand{\FigCap}[1]{\footnotesize\par\noindent Fig.\  %
  \refstepcounter{figure}\thefigure. #1\par}

\newcommand{\TableFont}{\footnotesize}
\newcommand{\TableFontIt}{\ttit}
\newcommand{\SetTableFont}[1]{\renewcommand{\TableFont}{#1}}
  
\newcommand{\MakeTable}[4]{\begin{table}[htb]\TabCap{#2}{#3}
  \begin{center} \TableFont \begin{tabular}{#1} #4   
  \end{tabular}\end{center}\end{table}}

\newcommand{\MakeTableSep}[4]{\begin{table}[p]\TabCap{#2}{#3}
  \begin{center} \TableFont \begin{tabular}{#1} #4
  \end{tabular}\end{center}\end{table}}
\newcommand{\TabCapp}[2]{\begin{center}\parbox[t]{#1}{\centerline{
  \small {\spaceskip 2pt plus 1pt minus 1pt T a b l e}
  \refstepcounter{table}\thetable}
  \vskip2mm
  \centerline{\footnotesize #2}}
  \vskip3mm
\end{center}}

\newcommand{\MakeTableSepp}[4]{\begin{table}[p]\TabCapp{#2}{#3}\vspace*{-.7cm}

  \begin{center} \TableFont \begin{tabular}{#1} #4
  \end{tabular}\end{center}\end{table}}

\newfont{\bb}{ptmbi8t at 12pt}
\newfont{\bbb}{cmbxti10}
\newfont{\bbbb}{cmbxti10 at 9pt}
\newcommand{\uprule}{\rule{0pt}{2.5ex}}
\newcommand{\douprule}{\rule[-2ex]{0pt}{4.5ex}}
\newcommand{\dorule}{\rule[-2ex]{0pt}{2ex}}
\def\thefootnote{\fnsymbol{footnote}}

\newenvironment{references}%
{
\footnotesize \frenchspacing
\renewcommand{\thesection}{}
\renewcommand{\in}{{\rm in }}
\renewcommand{\AA}{Astron.\ Astrophys.}
\newcommand{\AAS}{Astron.~Astrophys.~Suppl.~Ser.}
\newcommand{\ApJ}{Astrophys.\ J.}
\newcommand{\ApJS}{Astrophys.\ J.~Suppl.~Ser.}
\newcommand{\ApJL}{Astrophys.\ J.~Letters}
\newcommand{\AJ}{Astron.\ J.}
\newcommand{\IBVS}{IBVS}
\newcommand{\PASP}{P.A.S.P.}
\newcommand{\Acta}{Acta Astron.}
\newcommand{\MNRAS}{MNRAS}
\renewcommand{\and}{{\rm and }}
\section{{\rm REFERENCES}}
\sloppy \hyphenpenalty10000
\begin{list}{}{\leftmargin1cm\listparindent-1cm   
\itemindent\listparindent\parsep0pt\itemsep0pt}}%
{\end{list}\vspace{2mm}}

\def\TYLDA{~}
\newlength{\DW}
\settowidth{\DW}{0}
\newcommand{\dw}{\hspace{\DW}}

\newcommand{\refitem}[5]{\item[]{#1} #2%
\def\REFARG{#3}\ifx\REFARG\TYLDA\else, {\it#3}\fi
\def\REFARG{#4}\ifx\REFARG\TYLDA\else, {\bf#4}\fi
\def\REFARG{#5}\ifx\REFARG\TYLDA\else, {#5}\fi.}
\newcommand{\Section}[1]{\section{\hskip-6mm.\hskip3mm#1}}
\newcommand{\Subsection}[1]{\subsection{#1}}
\newcommand{\Acknow}[1]{\par\vspace{5mm}{\bf Acknowledgements.} #1}
\pagestyle{myheadings}

\newcommand{\xrule}{\rule{0pt}{2.5ex}}
\newcommand{\xxrule}{\rule[-1.8ex]{0pt}{4.5ex}}
\def\thefootnote{\fnsymbol{footnote}}
\begin{center}
{\Large\bf The Optical Gravitational Lensing Experiment.\\
\vskip3pt
Catalog of RR Lyrae Stars\\
\vskip6pt
in the Large Magellanic Cloud\footnote{Based on
observations obtained with the 1.3~m Warsaw telescope at the Las
Campanas
Observatory of the Carnegie Institution of Washington.}}
\vskip1.2cm
{\bf I.~~S~o~s~z~y~\'n~s~k~i$^1$,~~A.~~U~d~a~l~s~k~i$^1$,~~M.~~S~z~y~m~a~{\'n}~s~k~i$^1$,\\
M.~~K~u~b~i~a~k$^1$,~~G.~~P~i~e~t~r~z~y~\'n~s~k~i$^{1,2}$,~~
P.~~W~o~\'z~n~i~a~k$^3$,\\ K.~\.Z~e~b~r~u~\'n$^1$,
~O.~~S~z~e~w~c~z~y~k$^1$~ and ~\L.~~W~y~r~z~y~k~o~w~s~k~i$^1$}
\vskip8mm
{$^1$Warsaw University Observatory, Al.~Ujazdowskie~4, 00-478~Warszawa, Poland\\
e-mail: (soszynsk,udalski,msz,mk,pietrzyn,zebrun,szewczyk,wyrzykow)@astrouw.edu.pl\\
$^2$ Universidad de Concepci{\'o}n, Departamento de Fisica,
Casilla 160--C, Concepci{\'o}n, Chile\\
$^3$ Los Alamos National Laboratory, MS-D436, Los Alamos, NM 87545 USA\\
e-mail: wozniak@lanl.gov}
\end{center}

\vskip0.6cm

\Abstract{We present the catalog of RR~Lyr stars discovered in a 4.5 square 
degrees area in the central parts of the Large Magellanic Cloud (LMC). 
Presented sample contains 7612 objects, including 5455 fundamental mode 
pulsators (RRab), 1655 first-overtone (RRc), 272 second-overtone (RRe) and 230 
double-mode RR~Lyr stars (RRd). Additionally we attach a~list of several dozen 
other short-period pulsating variables. The catalog data include astrometry, 
periods, {\it BVI} photometry, amplitudes, and parameters of the Fourier 
decomposition of the {\it I}-band light curve of each object. 

We present density map of RR~Lyr stars in the observed fields which shows that 
the variables are strongly concentrated toward the LMC center. The modal 
values of the period distribution for RRab, RRc and RRe stars are 0.573, 0.339 
and 0.276 days, respectively. The period--luminosity diagrams for {\it BVI} 
magnitudes and for extinction insensitive index $W_I$ are constructed. We 
provide the $\log P$--$I$, $\log P$--$V$ and $\log P$--$W_I$ relations for 
RRab, RRc and RRe stars. The mean observed {\it V}-band magnitudes of RR~Lyr 
stars in the LMC are 19.36~mag and 19.31~mag for ab and c types, respectively, 
while the extinction free values are 18.91~mag and 18.89~mag. 

We  found a large number of RR~Lyr stars pulsating in two modes closely 
spaced in the power spectrum. These stars are believed to exhibit non-radial 
pulsating modes. We discovered three stars which simultaneously reveal 
RR~Lyr-type and eclipsing-type variability. If any of these objects were an 
eclipsing binary system containing RR~Lyr star, then for the first time the 
direct determination of the mass of RR~Lyr variable would be possible. 

We provide a list of six LMC star clusters which contain RR~Lyr stars. 
The richest cluster, NGC~1835, hosts 84 RR~Lyr variables. The period 
distribution of these stars suggests that NGC~1835 shares features of 
Oosterhoff type I and type II groups. 

All presented data, including individual {\it BVI} observations and finding 
charts are available from the OGLE {\sc Internet} archive.}{Stars:
variables: RR Lyr -- Magellanic Clouds -- Catalogs} 

\Section{Introduction}
RR~Lyr stars have been recognized as excellent tracers of the oldest stellar 
populations, as well as distance indicators of the Galactic and extragalactic 
stellar systems. These stars are among the most prominent members of 
Population II objects, easy to identify due to their characteristic pulsation 
periods, shape of the light curves, luminosities and colors. 

The Large Magellanic Cloud (LMC), as the nearest non-dwarf galaxy, plays a key 
role in studying a wide variety of astrophysical problems, because one can 
assume that the stars in the LMC are located, in the first approximation, at 
the same, relatively small distance. Since the population of RR~Lyr variables 
in the LMC is extremely rich, many important problems, including the 
distance scale, structural and chemical evolution of the Local Group galaxies, 
features and properties of the old-age stars may be studied with these stars. 

The first major survey for RR~Lyr stars in the LMC was performed by Thackeray 
and collaborators (Thackeray and Wesselink 1953). They found variables in the 
clusters NGC~1466 and NGC~1878. Surveys for field RR~Lyr stars in the 
Magellanic Clouds were initiated by Graham (1977). Sixty eight RR~Lyr 
variables were discovered in a ${1\arcd\times1\zdot\arcd3}$ field around the 
globular cluster NGC~1783. The subsequent surveys for the LMC RR~Lyr stars 
(Kinman \etal 1976, Nemec \etal 1985, Walker 1992, Hazen and Nemec 1992) 
increased the number of known RR~Lyr variables mainly in the outer fields of 
the LMC where density of stars is lower than in the central parts. 

The crucial moment in studies of RR~Lyr stars from the LMC was announcement of 
discovery of about 7900 such stars in 22 fields in the bar region of the LMC 
by the MACHO microlensing group (Alcock \etal 1996). In following papers the 
MACHO team presented discovery of 181 double-mode RR~Lyr variables (Alcock 
\etal 1997, 2000), and detected non-radial pulsations in the first-overtone 
RR~Lyr stars (Alcock \etal 2000). Small scale survey for RR~Lyr (about 100 
objects) was also presented by Clementini \etal (2003). 

The Optical Gravitational Lensing Experiment (OGLE; Udalski, Kubiak and 
Szyma{\'n}ski 1997) is the long term large microlensing survey  
started in the early 1990s. The LMC was included to the list 
of observed targets of the OGLE survey at the beginning of the second phase of 
the project (OGLE-II), in January 1997. The galaxy was regularly monitored to 
the end of the OGLE-II phase, \ie to November 2000. The LMC is also one of the 
main targets of the third phase of the project (OGLE-III), which started in 
June 2001. 

The OGLE-II survey substantially increased the number of known variable stars 
in both Magellanic Clouds. In the previous papers we presented the catalogs of 
eclipsing binary stars in the Magellanic Clouds (Udalski \etal 1998b, 
Wyrzykowski \etal 2003), catalogs of Cepheids in the Magellanic Clouds 
(Udalski \etal 1999ab), catalog of RR~Lyr stars in the SMC (Soszy{\'n}ski 
\etal 2002, hereafter Paper I) and general catalog of variable stars detected 
in the Magellanic Clouds ({\.Z}ebru{\'n} \etal 2001). In addition, {\it BVI} 
maps of the LMC and SMC were released providing precise photometry and 
astrometry of about 7 and 2~million stars from these galaxies, respectively 
(Udalski \etal 1998a, 2000). 

In this paper we present the catalog of 7612 RR~Lyr stars and several other 
pulsating objects, likely $\delta$~Sct stars, detected in the OGLE-II fields 
in the LMC. The stars were selected from the reprocessed OGLE-II photometry 
based on the Difference Image Analysis (DIA) technique. We provide all basic 
observational parameters of detected stars and present statistics of the most 
important parameters. The photometry and finding charts of all detected 
objects are available to the astronomical community from the OGLE {\sc 
Internet} archive. 

\vskip1cm
\Section{Observations and Data Reductions}
All observations presented in this paper were carried out during the second 
phase of the OGLE microlensing search with the 1.3-m Warsaw telescope at Las 
Campanas Observatory, Chile. The observatory is operated by the Carnegie 
Institution of Washington. The telescope was equipped with the ``first 
generation'' camera with a~SITe ${2048\times2048}$ CCD detector working in 
drift-scan mode. The pixel size was 24~$\mu$m giving the 0.417~arcsec/pixel 
scale. Observations of the LMC were performed in the ``slow'' reading mode of 
the CCD detector with the gain 3.8~e$^-$/ADU and readout noise of about 
5.4~e$^-$. Details of the instrumentation setup can be found in Udalski, 
Kubiak and Szyma{\'n}ski (1997). 

Observations presented in this paper were conducted between January 6, 1997 
and November 26, 2000. Eleven driftscan fields (LMC$\_$SC1--LMC$\_$SC10 and 
LMC$\_$SC12), each covering $14\zdot\arcm2 \times57\arcm$ on the sky, were 
observed since January 1997. Additional ten fields (LMC$\_$SC11 and 
LMC$\_$SC13--LMC$\_$SC21), added in October 1997, increased the total observed 
area of the LMC to about 4.5~square degrees. Photometry was obtained with the 
$BVI$ filters, closely resembling the standard system. Due to microlensing 
search observing strategy the majority of frames were taken in the $I$ 
photometric band (about 260--520 epochs depending  on the field). The 
remaining images were collected through the {\it V}-band (typically about 
30--70 epochs) and {\it B}-band (about 30 epochs) filters. The effective  
exposure time lasted 125, 174 and~237~seconds for the $I$, $V$ and 
{\it B}-band, respectively. The median seeing was about 1\zdot\arcs3 for our 
dataset. 

The analysis of the dataset was performed according to the procedures 
described in Paper I. The {\it I}-band photometry was obtained using 
Difference Image Analysis (DIA) -- image subtraction algorithm developed by 
Alard and Lupton (1998) and Alard (2000), and implemented by Wo{\'z}niak 
(2000). The DIA photometry pipeline was rerun on the complete set of OGLE-II 
images and the code included several modifications compared to the first 
application of the method (\.Zebru\'n \etal 2001). We found that the detection 
efficiency of faint variable objects (in particular RR~Lyr stars) in the 
\.Zebru\'n \etal (2001) catalog was relatively poor. 

The details of the DIA analysis and calibration of photometry may be found in 
Wo{\'z}niak (2000) and {\.Z}ebru{\'n}, Soszy{\'n}ski and Wo{\'z}niak (2001). 
The {\sc DoPhot} photometry package (Schechter, Saha and Mateo 1993) was used 
to determine $V$ and $B$-band magnitudes of the stars. For more details about 
data reduction and transformation procedures the reader is referred to Udalski 
\etal (2000). 

Equatorial coordinates of all stars were calculated in the identical manner as 
described in Udalski \etal (2000). The internal accuracy of the determined 
equatorial coordinates, as measured in the overlapping regions of neighboring 
fields, is about 0\zdot\arcs15--0\zdot\arcs20 with possible systematic errors 
of the DSS coordinate system up to 0\zdot\arcs7. 

\begin{figure}[htb]
\vspace{11cm}
\vskip5mm
\FigCap{Density map of RR~Lyr stars in the LMC. White circles indicate 
positions of star clusters where RR~Lyr stars were found. The cross marks 
probable center of the LMC.} 
\end{figure}
Fig.~1 presents density map of the RR~Lyr variables from the LMC. One can 
notice that, contrary to the RR~Lyr stars in the SMC (Paper~I), RR~Lyr 
variables in the LMC are strongly concentrated toward the galaxy center. The 
highest density region and approximately the center of the presented 
distribution has the equatorial coordinates RA=5\uph22\zdot\upm9 DEC=$-
69\arcd39\arcm$. It is possible that this is the center of the LMC. 

\vskip1cm
\Section{Interstellar Reddening}
The reddening map of the OGLE-II fields in the LMC was derived by Udalski 
\etal (1999a). Since the reddening toward the LMC is clumpy and variable, the 
mean reddening values were determined in 84 lines-of-sight -- four per each 
OGLE-II field. Udalski \etal (1999a) used red clump stars for mapping the 
fluctuations of the mean reddening in the LMC, taking their mean {\it I}-band 
magnitude as the reference brightness. Differences of the observed mean {\it 
I}-band magnitudes were assumed as differences of the mean $A_I$ extinction, 
and then converted to differences of $E(B-V)$. The zero points of the 
reddening map were derived based on previous determinations of reddening 
around clusters NGC~1850, NGC~1835, and in the field of the eclipsing variable 
star HV2274. 

\MakeTable{lcccc}{12.5cm}{$E(B-V)$ reddening in the LMC fields}
{\hline
\multicolumn{1}{c}{Field}& Subfield~1   & Subfield~2  & Subfield~3  & Subfield~4\\
            & $E(B-V)$     & $E(B-V)$    & $E(B-V)$    & $E(B-V)$\\
\hline
LMC$\_$SC1 & 0.117 & 0.152 & 0.147 & 0.163 \\ LMC$\_$SC2 & 0.121 & 0.121 &
0.150 & 0.131 \\ LMC$\_$SC3 & 0.134 & 0.120 & 0.123 & 0.117 \\ LMC$\_$SC4 &
0.130 & 0.120 & 0.105 & 0.118 \\ LMC$\_$SC5 & 0.130 & 0.115 & 0.108 & 0.133
\\ LMC$\_$SC6 & 0.138 & 0.125 & 0.107 & 0.123 \\ LMC$\_$SC7 & 0.143 & 0.138
& 0.142 & 0.146 \\ LMC$\_$SC8 & 0.131 & 0.133 & 0.136 & 0.142 \\ LMC$\_$SC9
& 0.143 & 0.165 & 0.156 & 0.149 \\ LMC$\_$SC10& 0.156 & 0.147 & 0.146 &
0.132 \\ LMC$\_$SC11& 0.147 & 0.154 & 0.150 & 0.152 \\ LMC$\_$SC12& 0.152 &
0.146 & 0.127 & 0.139 \\ LMC$\_$SC13& 0.154 & 0.129 & 0.135 & 0.130 \\
LMC$\_$SC14& 0.124 & 0.142 & 0.138 & 0.127 \\ LMC$\_$SC15& 0.145 & 0.125 &
0.147 & 0.126 \\ LMC$\_$SC16& 0.135 & 0.148 & 0.185 & 0.181 \\ LMC$\_$SC17&
0.171 & 0.193 & 0.175 & 0.201 \\ LMC$\_$SC18& 0.182 & 0.178 & 0.173 & 0.170
\\ LMC$\_$SC19& 0.153 & 0.153 & 0.187 & 0.167 \\ LMC$\_$SC20& 0.132 & 0.137
& 0.142 & 0.163 \\ LMC$\_$SC21& 0.133 & 0.152 & 0.145 & 0.146 \\
\hline}

The final $E(B-V)$ reddening in our LMC fields is listed in Table~1. The error 
of the map is estimated to be $\pm0.02$ mag. Interstellar extinction in the 
{\it BVI} bands can be calculated using the standard extinction curve 
coefficients (\eg Schlegel \etal 1998): 
$$A_B=4.32\cdot E(B-V),\quad A_V=3.24\cdot E(B-V),\quad A_I=1.96\cdot E(B-V)$$ 

\Section{Selection of RR Lyr Stars}
The search for variable objects in the LMC fields was performed using 
observations in the {\it I}-band as they are much more numerous than in other 
bands. First, a preliminary search for RR~Lyr variables was performed using 
the regular OGLE-II PSF ({\sc DoPhot}) photometry. Candidates for variable 
stars were selected based on comparison of the standard deviation of all 
individual measurements of a~star with typical standard deviation for stars of 
similar brightness. Light curves of selected candidates were searched for 
periodicity using the {\sc AoV} algorithm (Schwarzenberg-Czerny 1989). Light 
curves of all objects revealing statistically significant periodic signal were 
then visually inspected and divided into a~few groups of variable stars. 

In the second stage we checked variability of stars using their DIA 
photometry. We selected stars with the mean {\it I}-band magnitude in the 
range 15~mag to 20~mag, and with the standard deviation of photometry at least 
0.01~mag bigger than the median value of standard deviation for stars at given 
brightness. Photometry of each star from this list was subject to a 
period-searching program {\sc Fnpeaks} (Ko{\l}aczkowski 2003 -- private 
communication). This algorithm is relatively fast what made it possible to 
test about 2 million stars. The output from {\sc Fnpeaks} contains the most 
significant peaks from the power spectrum (we checked three most significant 
frequencies of all stars) with a signal-to-noise parameter characterizing each 
peak. Then, we performed Fourier analysis of stars with periods shorter than 1 
day, and signal-to-noise higher than 3.5, \ie corresponding to the boundary 
between constant and variable stars. For further analysis we selected objects 
which occupy appropriate region in the $\log P$--$R_{21}$ diagram (see 
Section 8.4). In the final step, each object was carefully checked visually. 
Other type of variable stars and obvious artifacts were removed from the 
sample. 

While the vast majority of non-pulsating variable stars were removed from the 
catalog in the selection process, it is possible that some of the eclipsing 
binary stars could remain on the list of RRc stars, especially when they have 
similar luminosity and colors like RR~Lyr stars. It is also likely that 
a limited number of other type of short-period pulsators, like $\delta$~Sct 
stars, anomalous Cepheids, pulsating blue stragglers, etc. are hidden among 
our sample of RR~Lyr object candidates. 

\vspace*{3mm}
\Section{Classification}
\Subsection{Single-Mode RR Lyr stars}
All selected variable star candidates were divided into five groups: 
fundamental mode RR~Lyr stars (RRab), first overtone (RRc), second overtone 
(RRe), double mode RR~Lyr stars (RRd) and other variables. 

The primary criterion of classification of single-mode RR~Lyr stars was their 
location in the $\log P$--$R_{21}$ diagram (see Section 8.4). When the errors 
of the Fourier parameter $R_{21}$ were large and in doubtful cases we also 
used the $\log P$--amplitude diagram (see Section 8.3). Fundamental mode and 
first-overtone pulsators form well-separated groups in both diagrams, so that 
selection of these classes of RR~Lyr stars was not difficult. 

The second-overtone RR~Lyr stars are thought to form a secondary peak in the 
period distribution of overtone RR~Lyr variables. Such a peak, around 
${P=0.281}$ days, was first discovered and interpreted by Alcock \etal (1996). 
Studying $\log P$--$R_{21}$ and $\log P$--amplitude diagrams we noticed that 
stars with period smaller than about 0.3 days can be divided into two groups: 
sinusoidal and low amplitude pulsators and more asymmetric higher amplitude 
RR~Lyr stars. Similarly to Clement and Rowe (2000), who studied OGLE 
photometry of RR~Lyr stars from the globular cluster $\omega$~Cen, we assumed 
that the first group is formed by RRe, \ie second-overtone stars. 

However, one should remember that other interpretations of these short-period 
low-amplitude stars are also possible. For example, models of Bono \etal 
(1997) indicate that period--amplitude sequence of RRc stars has a 
characteristic ``bell'' shape, \ie the first-overtone pulsators with shortest 
periods should have lower amplitudes. Kov{\'a}cs (1998) presented aguments 
that this short-period class of RR~Lyr variables are first-overtone pulsators. 
He used pulsation models and evolutionary calculations to show that the 
second-overtone oscilations with a period as long as 0.28~days are not 
possible. Moreover, Stellingwerf, Gautschy and Dickens (1987) theoretically 
predicted that the second-overtone RR~Lyr stars should have sharper peak at 
maximum of light than first-overtone pulsators. We do not find such effect in 
our observational data -- light curves of our RRe candidates are more 
symmetrical than light curves of RRc stars at the same periods. 

In contrast to the fundamental mode and first-overtone RR~Lyr stars, the 
observational parameters of RRc and RRe variables overlap, thus it is 
difficult to make definitive classification. We stress that our division into 
first- and second-overtone RR~Lyr stars has only statistical sense and in 
individual cases might be wrong. 

\vspace*{2mm}
\Subsection{Multi-Mode RR Lyr stars}
Double-mode RR~Lyr stars (RRd) are relatively easy to detect, because of
their well defined range of the period ratio. First, the preliminary search
for RRd stars was carried out using the output of {\sc Fnpeaks}. We
selected for visual inspection objects for which the ratio of the two of
three highest peaks in the power spectrum was close to 0.745. The second
search for double-mode RR~Lyr stars was performed by fitting a fourth order
Fourier series to each light curve from our sample and subtracting fitted
function from the observational data. Then, the residuals were searched for
other periodic signals and, if detected, such a~candidate was marked for
visual inspection.

During the process of selection of double-mode RR-Lyr variables we discovered 
numerous class of multi-periodic stars which exhibit two frequencies, very 
closely spaced. We found that such a~phenomenon occurs in about 15\% of RRab, 
6\% of RRc and 23\% of RRe stars, although one should remember that we 
performed only preliminary frequency analysis. The range of period ratios of 
the secondary and primary periodicities depends on the RR~Lyr type: for RRab 
stars it is usually from 0.98 to 1.03, for RRc and RRe stars the period ratio 
ranges from 0.95 to 1.07, although smaller and larger ratios also sometimes 
occur. 

This class of multi-periodic RR~Lyr variables was described by Olech \etal 
(1999), who found such stars in the globular clusters M5 and M55. They argued 
that such a behavior is caused by the presence of non-radial oscillations in 
those stars. Alcock \etal (2000) performed frequency analysis of the variables 
from the LMC classified previously as RRc stars and divided multi-periodic 
pulsators into nine groups according to their frequency spectra. High period 
ratios in RR~Lyr stars were also detected in the OGLE Galactic bulge RR~Lyr 
sample (Moskalik 2000) and in RR~Lyr stars from the SMC (Paper I). Theoretical 
models of these stars were proposed by Van Hoolst, Dziembowski and Kawaler 
(1998), Dziembowski and Cassisi (1999), Nowakowski and Dziembowski (2001). An 
extensive and careful frequency analysis of objects from our catalog should 
allow to detect a large variety of multi-periodic and non-stationary RR~Lyr 
stars. 

We also found a few double-mode variables with periods and period ratios 
outside the region occupied by ``classical'' RRd stars. Probably four stars 
with the period ratio of about 0.76 and longer periods between 0.2 and 0.4 
days are double-mode $\delta$~Sct stars. Two similar stars were detected in 
the SMC (Paper~I). Additional four stars are double-mode pulsators with period 
ratios from 0.8028 to 0.8062 and the longer dominant periods between 0.3 and 
0.4 days. Three of these stars were discovered by MACHO (Alcock \etal 2000). 
Such ratio of periods is theoretically predicted as an indicator of RR~Lyr 
variables pulsating in the first and second overtone. Because these objects 
are about 1~mag brighter than typical RR~Lyr stars, and their light curves are 
different than light curves of RR~Lyr stars, we suggest that these are other 
type of variable stars, perhaps short period double-mode Cepheids. Further 
observations and studies of these objects would be necessary to confirm their 
status. 

\begin{figure}[htb]
\vspace*{-4mm}
\centerline{\includegraphics[width=13.5cm]{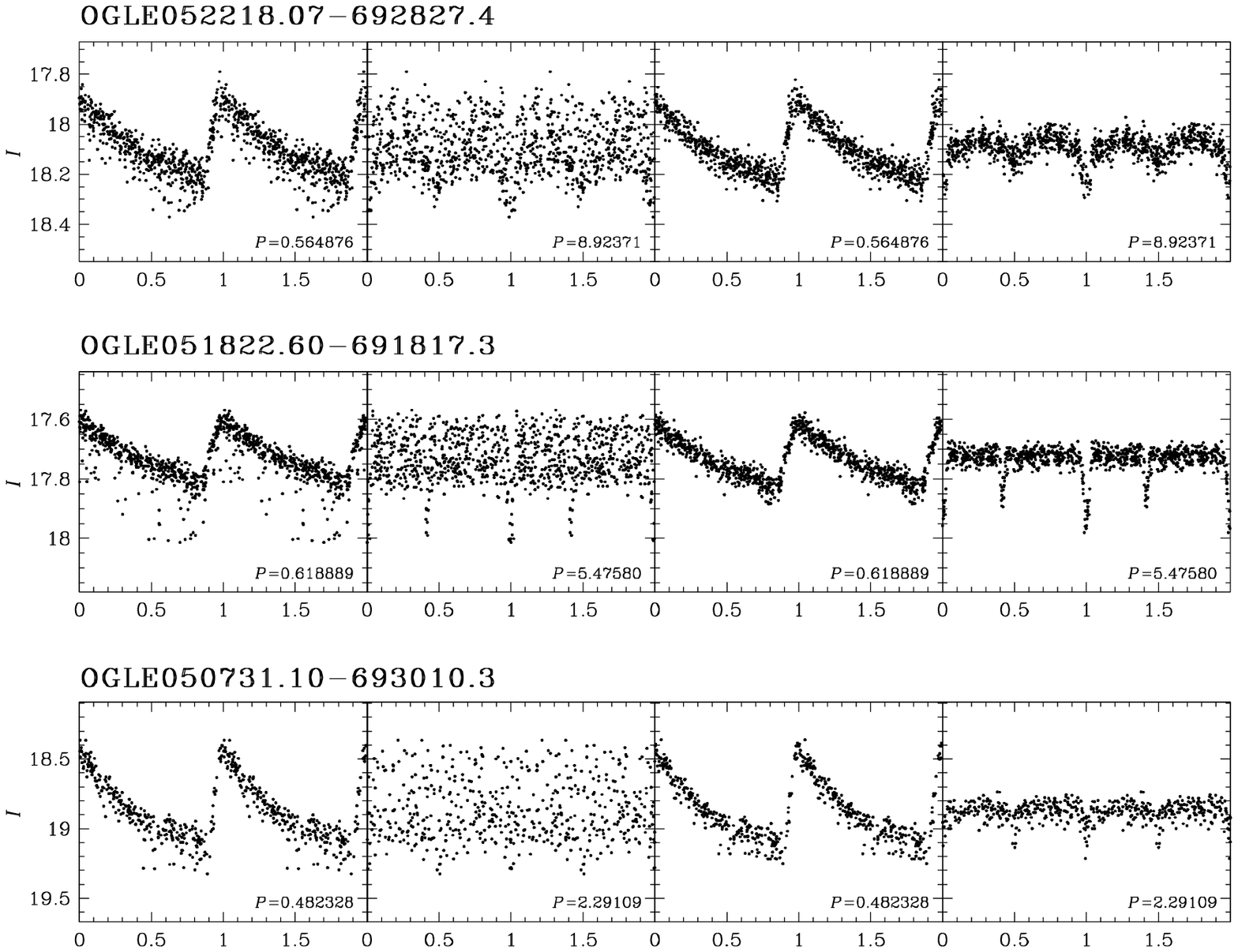}} 
\vspace*{-7.8cm}
\FigCap{Light curves of three RR~Lyr stars revealing also eclipsing 
variability. In the first two columns the original photometric data folded 
with the RR~Lyr period and eclipsing period are presented. In the third and 
the fourth columns RR~Lyr and eclipsing light curves after subtracting the 
other component are shown.} 
\end{figure} 
Finally, we detected three objects exhibiting simultaneously the RR~Lyr and 
eclipsing binary star type of variability. Their light curves are presented in 
Fig.~2. In the first two columns original light curves folded with pulsation 
and eclipsing periods are presented. The third and fourth columns show the 
light curve of  RR~Lyr and eclipsing star after subtracting the other 
variability. 

The stars can be either optical blends of a~RR~Lyr star with physically 
unrelated eclipsing system unresolved within the seeing disk or eclipsing 
systems containing RR~Lyr object as one of the components. The latter case 
would be extremely interesting and of great importance as the photometric and 
spectroscopic follow-up observations should allow for the first time to 
directly determine the mass and radius of RR~Lyr star. 

Our preliminary analysis suggests that OGLE050731.10--693010.3 is probably an 
optical blend. The eclipsing period is too short for the system to be 
physically related: the typical RR~Lyr star would be larger than the 
corresponding Roche lobe. In the case of OGLE051822.60--691817.3 the situation 
is unclear. The RR~Lyr component could fit its Roche lobe but it should be 
close to its size. So the star should be tidally distorted. On the other hand 
the shape of the eclipsing light curve with flat parts between eclipses does 
not show any significant ellipsoidal effect. Thus OGLE052218.07--692827.4 
remains the best candidate for an eclipsing system containing RR~Lyr 
component. As all objects are within reach of accurate spectroscopy obtained 
with large 8-m class telescopes, radial velocity measurements should clarify 
the status of all these interesting objects. 

\vspace*{3mm}
\Section{Catalog of RR Lyr Stars}
5455 RRab, 1655 RRc, 230 RRd and 272 RRe variable stars passed our selection 
criteria. Due to large number of objects the entire catalog is available only 
electronically from the OGLE {\sc Internet} archive: 
\begin{center}
{\it http://ogle.astrouw.edu.pl} \\ 
{\it ftp://ftp.astrouw.edu.pl/ogle/ogle2/var\_stars/lmc/rrlyr/}\\
\end{center}
or its US mirror
\vspace*{-10pt}
\begin{center}
{\it http://bulge.princeton.edu/\~{}ogle}\\ {\it
ftp://bulge.princeton.edu/ogle/ogle2/var\_stars/lmc/rrlyr/}\\
\end{center}
\begin{figure}[p]
\vspace{15cm}
\vspace*{-2cm}
\FigCap{Exemplary light curves of RR~Lyr stars in the LMC. In the first two 
rows light curves of eight RRab stars arranged according to the periods are  
presented. In the next rows samples of RRc and RRe stars are presented. Bottom  
row shows the light curves of an exemplary RRd variable -- original  
photometric data folded with the shorter and longer periods, and light curves  
of each mode after subtraction of the other period variability.} 
\end{figure} 

\begin{figure}[p]
\psfig{figure=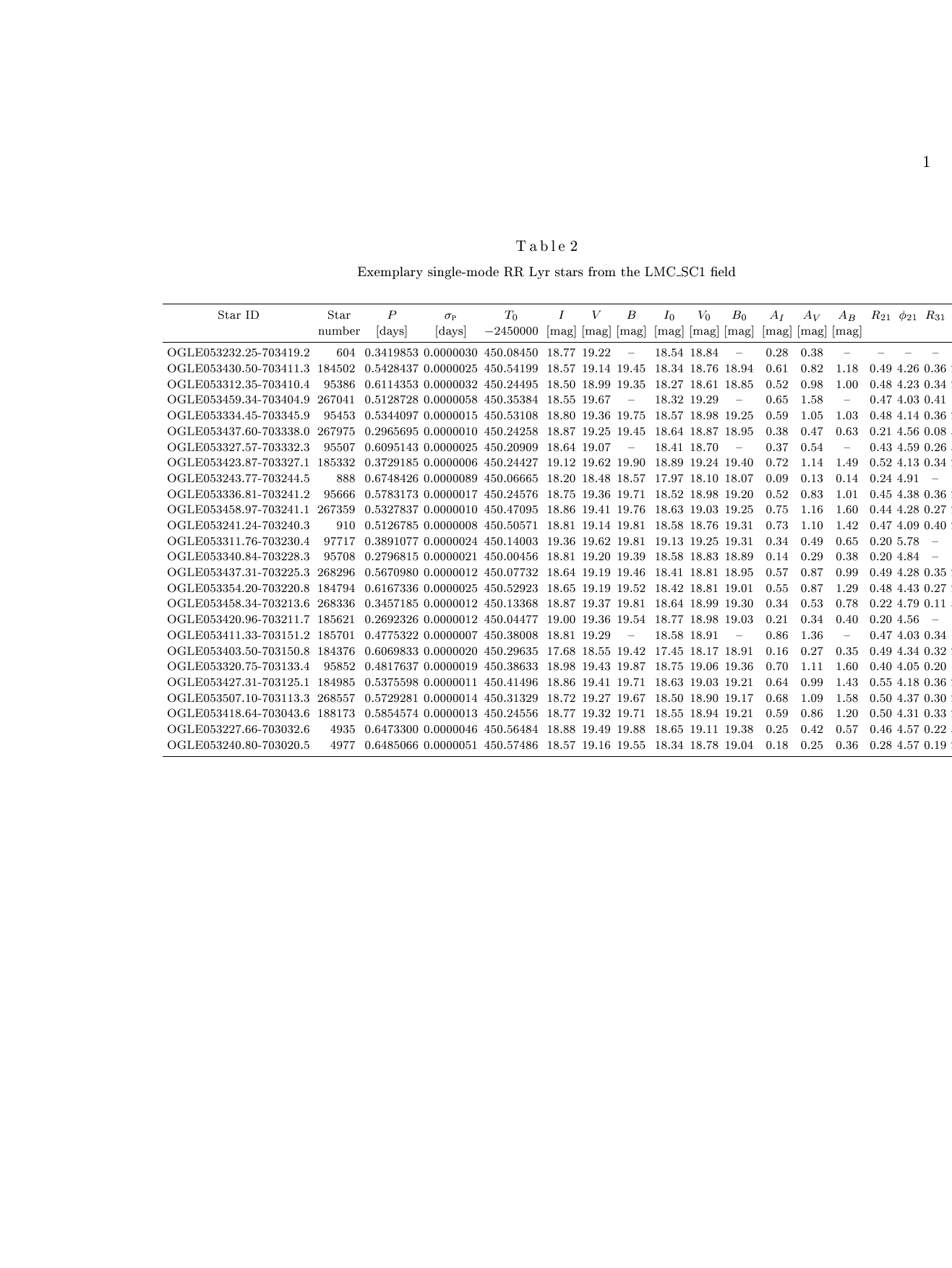,bbllx=0pt,bblly=0pt,bburx=670pt,bbury=685pt,angle=90,clip=}
\end{figure}

In tables we provide the following data for each object: star ID (which 
includes equatorial coordinates of the star, RA and DEC, for the epoch 
2000.0), star number (consistent with published LMC maps, Udalski \etal 2000), 
period in days, period error, moment of the zero phase corresponding to 
maximum light, intensity mean {\it IVB} photometry, extinction free 
magnitudes, amplitudes of the {\it IVB}-band light curves, Fourier parameters 
of the light curve decomposition and pulsation type. As an example we present 
in Table~2 the data for the first 27~single-mode RR~Lyr variables.

Individual {\it BVI} observations of all objects and finding charts are also 
available from the OGLE {\sc Internet} archive. For double-mode RR~Lyr stars 
we provide both periods, period ratios, and other parameters, separately for 
fundamental mode and first-overtone of each star. Additionally, we provide the 
lists of multi-periodic RR~Lyr stars probably pulsating in non-radial modes 
and objects found in star clusters in the LMC. 

The lists contain altogether 8117 entries but only 7612 objects, because 500 
RR~Lyr stars were detected twice or even three times in the overlapping 
regions between adjacent fields. One of the files in the archive contains 
cross-reference list to identify stars in the overlapping regions. 

Several exemplary light curves of all classes of RR~Lyr stars are presented in 
Fig.~3. The light curves of RRab, RRc and RRe stars are arranged according to 
periods. One should note that the diagrams have the same magnitude range to 
compare the amplitudes and brightness of stars. 

\vspace*{5mm}
\Section{RR Lyr Stars in the LMC Star Clusters}
Studies of RR~Lyr stars in star clusters may provide significant clues to 
several problems in the theory of pulsation and stellar evolution. 
Traditionally, RR~Lyr stars indicate a population of 11 or more Gyr old, 
although there is no strong evidence that can rule out their presence in 
substantially younger stellar systems. Therefore, detection of RR~Lyr 
variables in star clusters younger than about 10~Gyr would have direct 
consequences on our ideas about evolution of old stellar systems. Discovery of 
RR~Lyr stars in the substantial number of star clusters in the LMC may provide 
better understanding of the Oosterhoff dichotomy (Oosterhoff 1939). 

We used the OGLE catalog of the LMC star clusters (Pietrzy{\'n}ski \etal 1999) 
containing 745 clusters, their coordinates and angular sizes, to select RR~Lyr 
stars from the LMC clusters. We scanned our sample of RR~Lyr variables looking 
for stars in a~distance smaller than the cluster radius from the cluster 
center. 

Since the cores of some star clusters are extremely dense, we put additional 
effort for detecting RR~Lyr stars in those regions. For stars located near the 
center of the richest star clusters we lowered the limit for the 
signal-to-noise parameter of the light curve and then checked visually all new 
candidates. We improved in this way the completeness of our sample, but the 
completeness of our catalog is still the lowest in the cores of star clusters 
(see Section~9). We also stress that many stars in the central parts of the 
clusters are often severely blended. 

\hglue-3pt In 6 clusters, namely NGC~1835, NGC~1898, NGC~1916, NGC~1928, 
NGC~2005, and NGC~2019, we found more than 6 RR~Lyr stars. The results of our 
search are summarized in Table~3. 

\setcounter{table}{2}
\renewcommand{\TableFont}{\footnotesize}
\MakeTable{
c@{\hspace{8pt}} c@{\hspace{6pt}} c@{\hspace{6pt}} c@{\hspace{8pt}}
c@{\hspace{6pt}} c@{\hspace{6pt}} c@{\hspace{6pt}} c@{\hspace{6pt}}
c@{\hspace{6pt}} c@{\hspace{2pt}} }{12.5cm}{Star clusters containing RR~Lyr
stars}{\hline
\noalign{\vskip3pt}
Cluster & RA & Dec & Cluster & $N_{\rm RR}$ & $N_{\rm ab}$ & $N_{\rm c}$ & 
$N_{\rm e}$ & $N_{\rm d}$
& $N_{\rm field}$ \\ name & (J2000) & (J2000) & radius [\arcs] & & & & & &
(estimated) \\
\noalign{\vskip3pt}
\hline
\noalign{\vskip3pt}
NGC1835 & 5\uph05\upm07\ups & $-69\arcd24\arcm14\arcs$ & 70 & 84 & 55 &  21
& 2 & 6 & 2 \\ NGC1898 & 5\uph16\upm41\ups & $-69\arcd39\arcm24\arcs$ & 35
& 28 & 17 &  10 & 1 & 0 & 1 \\ NGC1916 & 5\uph18\upm38\ups &
$-69\arcd24\arcm23\arcs$ & 62 & 14 &  7 &   7 & 0 & 0 & 3 \\ NGC1928 &
5\uph20\upm58\ups & $-69\arcd28\arcm40\arcs$ & 31 &  7 &  6 &   1 & 0 & 0 &
1 \\ NGC2005 & 5\uph30\upm10\ups & $-69\arcd45\arcm10\arcs$ & 57 & 11 &  6
&   4 & 1 & 0 & 2 \\ NGC2019 & 5\uph31\upm56\ups & $-70\arcd09\arcm33\arcs$
& 49 & 41 & 24 &  13 & 2 & 2 & 1 \\
\noalign{\vskip3pt}
\hline}

In all these cases it is extremely unlikely that all detected RR~Lyr stars are 
simply field stars in the foreground or background of the clusters. To make 
sure that the majority of RR~Lyr stars are the cluster members, we estimated 
the number of field RR~Lyr variables, which should lie in the line-of-sight of 
the clusters. We counted RR~Lyr stars from our sample in the ring around each 
cluster. The inner radius of the ring was set to 100\arcs and the outer to 
400\arcs. Comparing angular field of the ring and cluster, we derived the 
expected number of field RR~Lyr stars inside the cluster area. These numbers 
are also provided in Table~3. They are significantly smaller than the number 
of detections. 

In several clusters we found one, two or three RR~Lyr stars. As the number of 
expected field RR~Lyr stars is similar we are unable to conclude whether these 
stars are real members of clusters, or they are near the line-of-sight of the 
clusters by an optical coincidence. 

\begin{figure}[htb]
\vspace*{-7mm}
\centerline{\includegraphics[width=10.7cm]{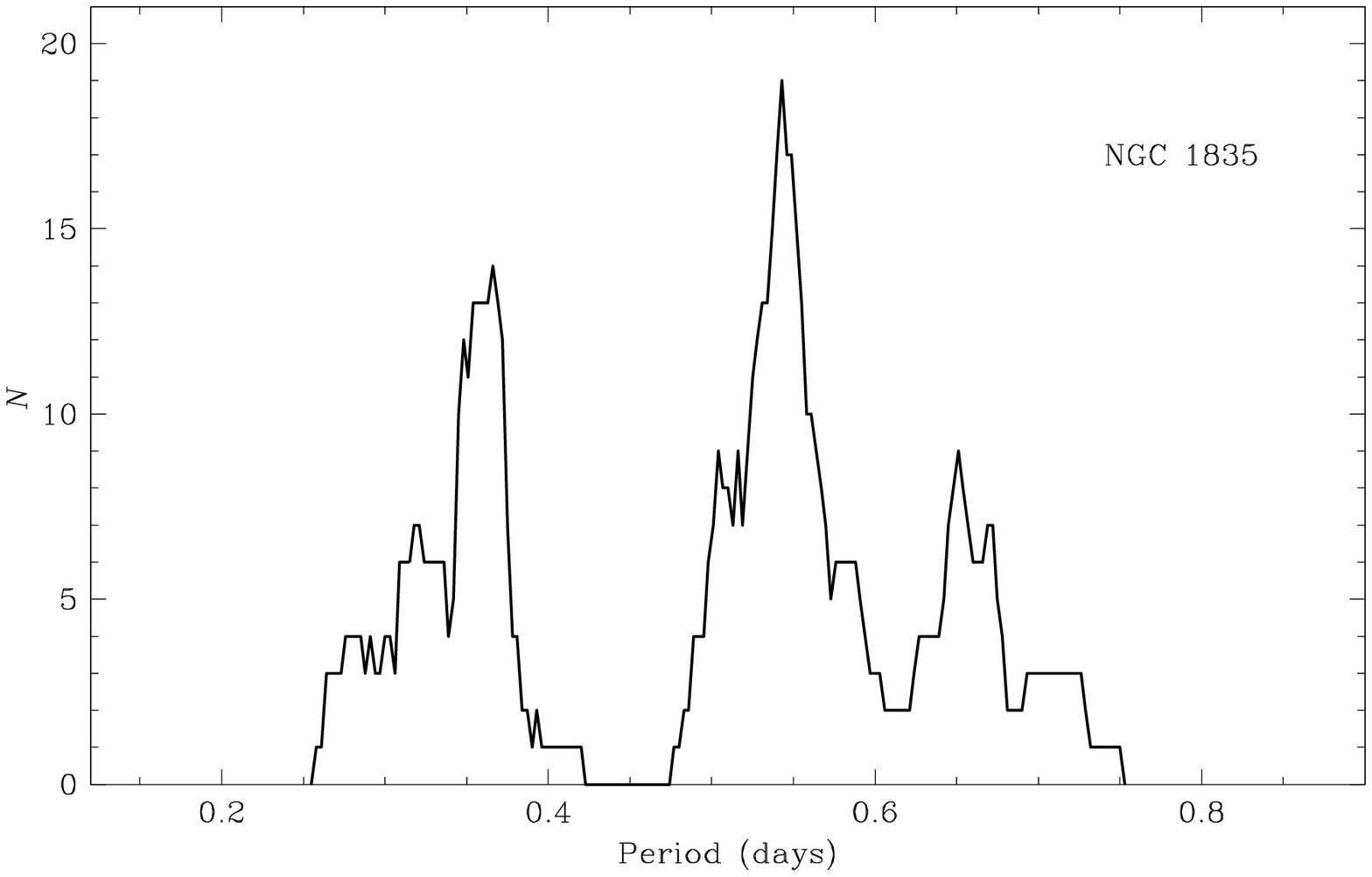}} 
\vspace*{-7.3cm}
\FigCap{Distribution of periods of 84 RR~Lyr stars from the star cluster 
NGC~1835. The function was obtained from ten histograms (with 0.03 days wide 
bins), shifted by 0.003 days to each other.} 
\end{figure}
NGC~1835 is the most compact star cluster in the observed region of the LMC. 
Graham and Ruiz (1974) and Walker (1992) surveyed this object for RR~Lyr 
variables and discovered in total 36 such stars. We detected 84 RR~Lyr 
variables: 55 RRab, 21 RRc, 6 RRd and 2 RRe stars. The period distribution of 
these stars is presented in Fig.~4. It is worth noticing that fundamental mode 
pulsators exhibit two peaks in the period histogram -- near 0.54 and 0.65 days. 
These peaks correspond to the well-known mean periods of the Oosterhoff type I 
(OoI) and Oosterhoff type II (OoII) clusters (Smith 1995). 

The second parameter which is used for distinction of the Oosterhoff's types 
is the ratio of the overtone RR~Lyr stars to the total number of RR~Lyr 
variables ($(N_{\rm c}+N_{\rm d}+N_{\rm e})/N_{\rm RR}$). For NGC~1835 this 
parameter is equal to 0.34, \ie between the typical values for OoI (0.17) and 
OoII (0.44). We conclude that NGC~1835 shares the features of both Oosterhoff 
groups. 

\vspace*{5mm}
\Section{Basic Parameters of RR Lyr Stars}
\Subsection{Period Distribution}
The period histogram is a widely used tool for studying RR~Lyr stars in 
galaxies and globular clusters because pulsational periods are unaffected 
neither by the distance nor by the interstellar reddening, and can be measured 
with the highest precision. Comparative analysis of the distributions of 
RR~Lyr variables from various stellar systems yields information about the 
early history of star formation. 

\begin{figure}[htb]
\vspace*{-7mm}
\centerline{\includegraphics[width=12cm]{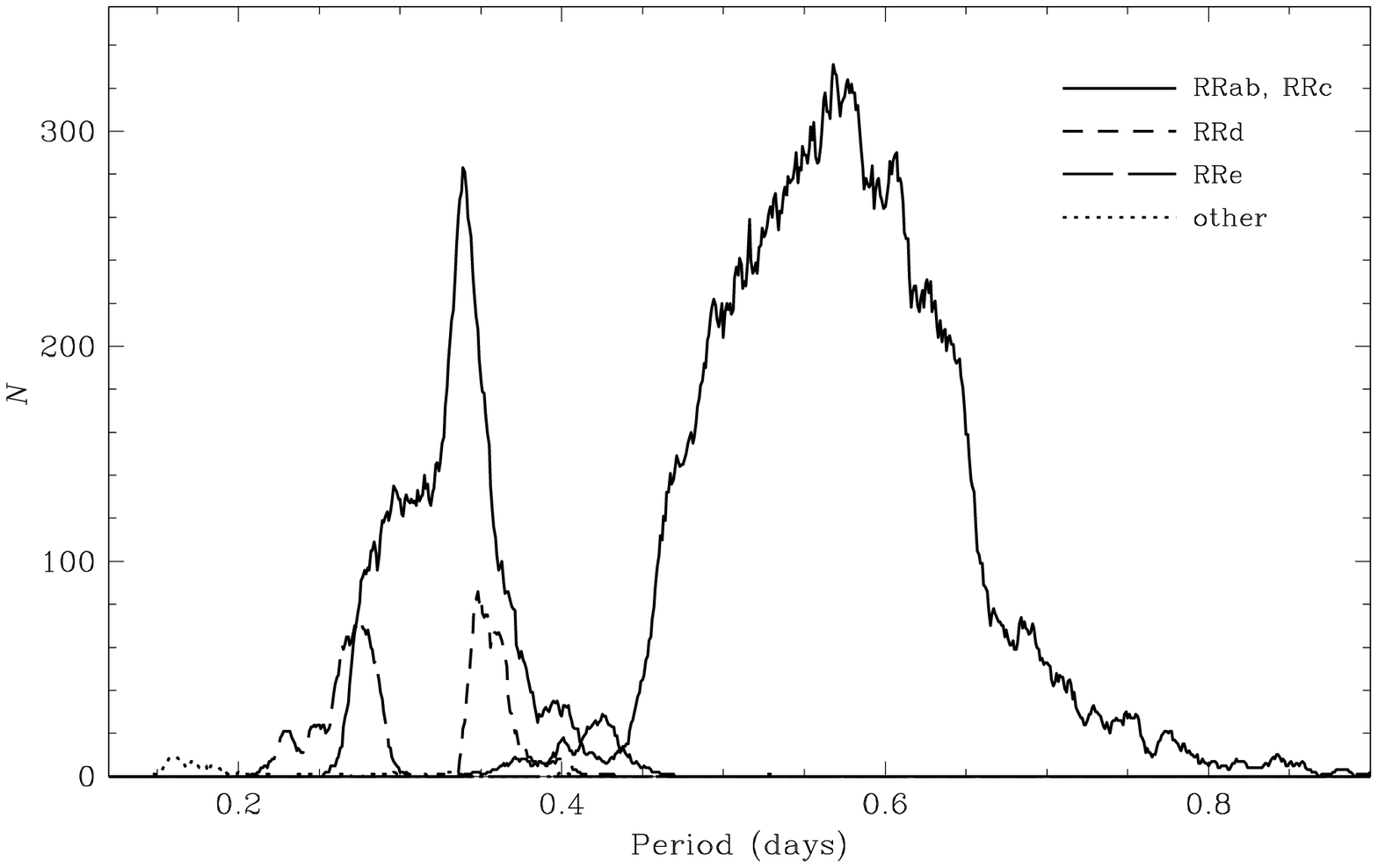}} 
\vspace*{-8.3cm}
\FigCap{Distribution of periods for RR~Lyr stars in the LMC. The function was 
obtained from ten histograms (with 0.01 days wide bins), shifted by 0.001 days 
to each other. Solid lines correspond to RRab and RRc stars, short-dashed line 
represents the first-overtone periods of RRd stars, long-dashed line -- RRe 
stars, and dotted line -- other stars.} 
\end{figure} 
The distribution of periods of 7612 RR~Lyr stars is presented in Fig.~5. The 
function is composed of series of ten histograms (bin width 0.01 days) 
shifted by 0.001 days with respect to each other. This manner of the period 
distribution presentation enables precise determination of the most likely 
periods of all types of the RR~Lyr stars. 

The mean period of ab-type RR~Lyr stars is ${\langle P_{ab}\rangle=0.573}$ 
days. This is a slightly lower value than that found by the MACHO team: 
${\langle P_{ab}\rangle=0.583}$ days (Alcock \etal 1996). The most likely 
period of RRc variables is equal to ${\langle P_c\rangle=0.339}$ days, what 
agrees well with the value obtained by MACHO (0.342~days). 

Alcock \etal (1996) also detected a secondary peak in the period distribution 
of the overtone pulsators. They interpreted it as due to presence of the 
second overtone RR~Lyr stars (RRe). Another explanation of the peak at 
${P=0.281}$ days was presented by Bono \etal (1997), who suggested that it 
could be a signature of a more metal-rich population of RR~Lyr stars. Similar 
secondary peaks were also detected in the Sculptor dwarf spheroidal galaxy 
(Ka{\l}u{\.z}ny \etal 1995), in the Oosterhoff type I and type II Galactic 
globular clusters (Clement \etal 2001) and in the Small Magellanic Cloud 
(Paper I). 

The secondary peak disappears when we separated first- and second-overtone 
RR~Lyr stars. In Fig.~5 the period distribution of objects classified by us 
as RRe stars is presented separately. The most likely period of these 
pulsators is ${\langle P_e\rangle=0.276}$ days, giving the mean 
${P_e/P_c=0.814}$. 

In Fig.~5 we also present the distribution of the first-overtone period of 
double-mode RR~Lyr stars. The function has a sharp maximum at ${P_d=0.348}$ 
days.

\vspace*{5mm}
\Subsection{Period-Luminosity Relations and Mean Magnitude}
Similarly to Paper I we derived the {\it BVI} intensity mean photometry for 
each object from our catalog. The light curves converted to the intensity 
units were approximated by the Fourier series of fifth order and integrated. 
Results were converted back to the magnitude scale. Accuracy of the mean {\it 
I}-band photometry is about 0.02~mag while that of the {\it V}-band and {\it 
B}-band about 0.05~mag and 0.08~mag, respectively, what is a consequence of 
smaller number of observations in the {\it BV}-bands. 

For each star we also determined the reddening free Wesenheit index (Madore 
1982), defined as: 
$$W_I=I-1.55(V-I)$$
where 1.55 is the ratio of total-to-selective absorption ($A_I/E(V-I)$).

\begin{figure}[p]
\vspace*{-9mm}
\vspace{15cm}
\vspace*{-7mm}
\FigCap{{\it IVB}-band period--luminosity diagrams for RR~Lyr stars in the 
LMC. Red points represent RRab stars, green -- RRc, blue -- RRd 
(first-overtone), magenta -- RRe, and black points -- other stars.} 
\end{figure} 

\begin{figure}[htb]
\vspace{6cm}
\FigCap{Period--luminosity diagram for extinction insensitive index ${W_I=I-
1.55(V-I)}$. Color of points represent the same RR~Lyr types as in Fig.~6.} 
\end{figure} 
Figs.~6 and 7 show the period--magnitude and period--$W_I$ diagrams, 
respectively. One can notice  that many stars are significantly brighter than 
typical RR~Lyr variables from the LMC while some are fainter. We left these 
stars on our list because they reveal all characteristics of the RR~Lyr-type 
light curves. Since the LMC fields are very crowded it is very likely that 
many of these variables are blended with other stars. Although quality of DIA 
photometry is much better compared to the standard profile photometry, it is 
impossible to separate blended stars, because the non-variable component of 
the photometry is measured with traditional methods in the reference image.  
In some cases, in particular when a measured star is located very close to 
other bright star, it is impossible to determine correct magnitudes. 

Moreover, other types of variable stars, \eg $\delta$~Sct and Anomalous 
Cepheids with similar light curves to RR~Lyr stars might be included to our 
catalog. They can be brighter or fainter than typical RR~Lyr star. Finally, 
there are RR~Lyr stars which are located in the halo of the LMC, thus, in 
front or behind the central parts of the galaxy, \ie brighter or fainter, 
respectively. 

Nevertheless, most of the RR~Lyr stars form distinct sequences in the 
period--luminosity diagrams. To derive period--luminosity relations we 
performed the following procedure for RRab and RRc stars, separately. First, 
we prepared histograms of the {\it BVI}-band magnitudes and $W_I$ index at 
several discrete values of $\log P$ (0.03 wide bins in $\log P$). In the next 
step we fitted a Gaussian to each of the histograms and determined maximum of 
the function, \ie we obtained the most likely brightness for a given period. 
Finally, we fitted a linear function to the $\log P $--magnitude points. 

We obtained strong period--luminosity relation for the {\it I}-band and rather 
weak correlation of $\log P $ and {\it V}-band mean magnitudes for both, 
ab-type and c-type RR~Lyr stars. We did not detect any statistically 
significant $\log P$--$B$ relation. Therefore we derived the most likely 
magnitude of the RR~Lyr stars in the {\it B}-band by fitting a Gaussian to the 
histograms prepared for the whole sample of RR~Lyr stars of a given type. 

The period--luminosity relations for RRab stars in the LMC are as follows:
\begin{eqnarray}
I&=&-1.62(\pm0.05)\log P+18.391(\pm0.011)\nonumber\\ 
V&=&-0.80(\pm0.08)\log P+19.150(\pm0.014)\nonumber\\ 
B&=&19.71(\pm0.03)\nonumber\\ 
W_I&=&-2.75(\pm0.04)\log P+17.217(\pm0.008)\nonumber
\end{eqnarray}
for RRc stars:
\vspace*{-21pt}
\begin{eqnarray}
I&=&-2.23(\pm0.03)\log P+17.840(\pm0.007)\nonumber\\ 
V&=&-1.03(\pm0.08)\log P+18.807(\pm0.012)\nonumber\\ 
B&=&19.58(\pm0.02)\nonumber\\ 
W_I&=&-2.88(\pm0.06)\log P+16.787(\pm0.015)\nonumber
\end{eqnarray}

Because the sample of RRe candidates is less numerous, we derived the
period--luminosity relations by fitting a~linear function to the mean 
magnitudes: 
\begin{eqnarray}
I&=&-3.24(\pm0.20)\log P+17.182(\pm0.012)\nonumber\\ 
V&=&-2.24(\pm0.20)\log P+18.156(\pm0.015)\nonumber\\ 
B&=&19.68(\pm0.04)\nonumber\\ 
W_I&=&-4.02(\pm0.08)\log P+16.337(\pm0.011)\nonumber
\end{eqnarray}

It is worth noticing that generally the slope of the period--luminosity 
relation depends on the type of RR~Lyr object. The dependence is steeper for 
stars pulsating in higher modes. 

Kov{\'a}cs and Walker (2001) derived period--luminosity--color relations for a 
sample of RR~Lyr stars of ab type from the Galactic globular clusters. They 
determined the dependence of period and quantity defined as ${X=V-2.5(V-I)}$, where 
$I$ and $V$ are magnitude average photometry. For comparison we calculated 
magnitude average luminosities for our sample of RRab stars and found 
$\log{P}$--$X$ relation. The slope obtained by Kov{\'a}cs and Walker (2001), 
${-2.51\pm0.07}$, is in very good agreement with our value: 
${-2.55\pm0.05}$. 

RR~Lyr stars are widely accepted as very good distance indicator and their 
mean magnitude in the {\it V}-band is often used for distance determinations. 
As we show in Fig.~6 the {\it V}-band magnitude might be somewhat dependent on 
the pulsation period, but the correlation is relatively weak in that band. 
Therefore we may determine the mean magnitudes of these stars in the LMC for 
future precise determination of the distance to this galaxy. The modal value 
of the distribution of the observed {\it V}-band magnitude of our sample of 
RR~Lyr is equal to ${19.36\pm0.03}$~mag and ${19.31\pm0.02}$~mag for ab and c 
types, respectively. For magnitudes dereddened with the OGLE-II reddening map 
(Section~3) the appropriate values are: ${18.91\pm0.02}$ and ${18.89\pm0.02}$, 
again for ab and c types, respectively. 

\vspace*{2mm}
\Subsection{Period-Amplitude Relations}
The period--amplitude diagram, known as Bailey diagram, is a widely used tool 
for analyzing features of RR~Lyr stars. Empirical and theoretical studies 
suggest that the distribution of RR~Lyr stars in the Bailey diagram depends on 
their metallicity. In Fig.~8 we plot the period--amplitude diagram for RR~Lyr 
stars from our catalog. 
\begin{figure}[p]
\vspace*{-1cm}
\vspace{15cm}
\vspace*{-1cm}
\FigCap{{\it IVB}-band period--amplitude diagrams for RR~Lyr stars in the LMC. 
Color of points represent the same RR~Lyr types as in Fig.~6.} 
\end{figure} 

Fundamental mode RR~Lyr stars present an anti-correlation between the period 
and amplitude. Due to large number of RRab stars one can notice that the 
sequence in the $\log P$--amplitude diagram is non-linear, what agrees with 
theoretical models (Bono \etal 1997). The width of the sequence is believed to 
be an effect of spread in the metal content. 

Distribution of the overtone pulsators in the Bailey diagram is more 
complicated. RRc and RRe stars present a weak anti-correlation of amplitudes 
and periods, and form continuous but overlapping sequences. The 
period--amplitude plot was an auxiliary tool to the $\log P$--$R_{21}$ 
diagram that was used for separation of the first-and second-overtone RR~Lyr 
stars. 

\vspace*{2mm}
\Subsection{Fourier Coefficients}
The {\it I}-band light curves of our RR~Lyr candidates were fitted to the 
Fourier series of the fifth order. Then, the Fourier parameters $R_{21}$, 
$R_{31}$, $\phi_{21}$ and $\phi_{31}$ were derived, where $R_{ij}=A_i/A_j$, 
$\phi_{ij}=\phi_{i}-i \phi_{j}$. $A_i$ and $\phi_{i}$ are the amplitude and 
phase terms of ($i-1$) harmonic of the Fourier decomposition of light curve. 

\begin{figure}[p]
\vspace*{-1cm}
\vspace{15cm}
\vspace*{-1cm}
\FigCap{$R_{21}$ and $\phi_{21}$ \vs $\log P$ diagrams for RR~Lyr stars in 
the LMC. Color of points represent the same RR~Lyr types as in Fig.~6.} 
\end{figure} 

The $R_{ij}$ and $\phi_{ij}$ parameters are widely used for analyzing 
properties of the light curves of pulsating stars. Using empirical and 
theoretical relations it is possible to derive physical characteristics of 
RR~Lyr stars, including mass, absolute magnitude, effective temperature and 
metallicity (\eg Kov{\'a}cs and Walker 2001). 

In Fig.~9 we plot $\log P$--$R_{21}$ and $\log{P}$--$\phi_{21}$ diagrams 
constructed for our sample of RR~Lyr stars. $\log P$--$R_{21}$ diagram is 
particularly  useful for dividing single-mode RR~Lyr stars into three classes: 
RRab, RRc and RRe, because these groups occupy different regions in the 
diagram. The $R_{21}$ parameter for each class of single-mode RR~Lyr variables 
tends to become smaller as the periods become longer. On the other hand the 
$\phi_{21}$ parameter of the RRab, RRc and RRe stars increases with periods. 
That means that the light curves of RR~Lyr with longer periods are more 
sinusoidal and more symmetrical than the light curves of stars with shorter 
periods pulsating in the same mode. 

Light curves of the first-overtone mode of double-mode pulsators seem to be 
on average more ``sharp'' than those of single mode stars with corresponding 
periods (\ie between 0.46 and 0.58 days). The fundamental mode pulsation of 
RRd stars has $R_{21}$ smaller than corresponding single-mode RR~Lyr 
variables, what means that the light curves are more sinusoidal than the light 
curves of RRab stars. 

\vspace*{2mm}
\Subsection{Petersen Diagram}
The period \vs period ratio diagram for double-mode pulsators, the so called 
Petersen diagram (Petersen 1973), is a powerful tool providing substantial 
information about the mass and metallicity of pulsating stars (Popielski, 
Dziembowski and Cassisi 2000). Since periods are firm and accurate 
observables, $P_0$~\vs~$P_1/P_0$ diagram together with theoretical models of 
double-mode RR~Lyr stars makes it possible to study various problems, 
including, for instance, the LMC distance scale (Kov{\'a}cs 2000). 
\begin{figure}[htb]
\vspace*{-1.1cm}
\centerline{\includegraphics[width=10cm]{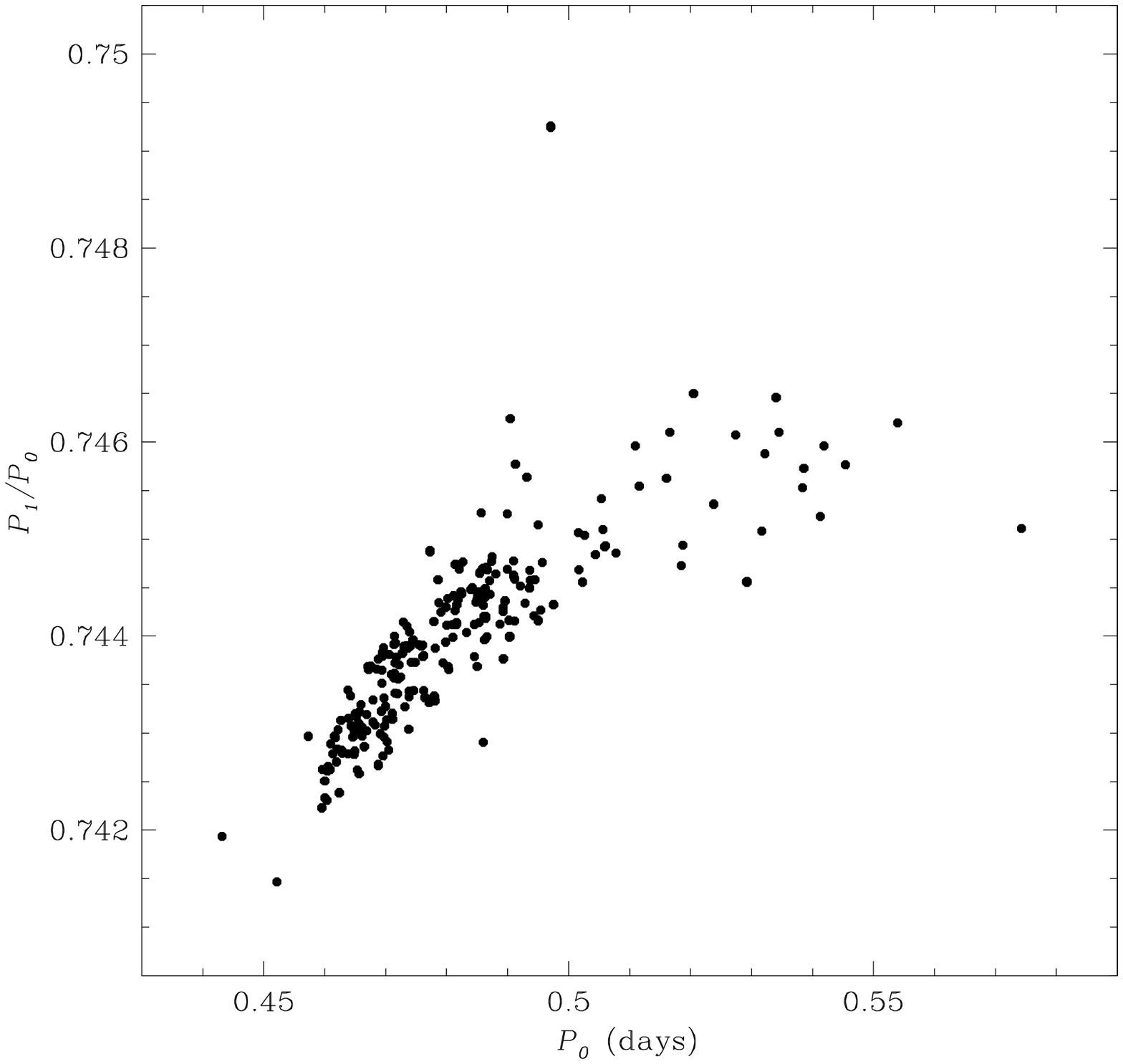}} 
\vspace*{-3.7cm}
\FigCap{Petersen diagram for the LMC RRd stars.} 
\end{figure} 

In Fig.~10 we show the Petersen diagram for 230 RRd stars from the LMC. 
Compared to double-mode RR~Lyr stars from the Galactic globular clusters and 
also from other galaxies of the Local Group, RRd variables in the LMC cover 
wide range of periods and period ratios, although most of stars concentrate 
near the shortest periods of the sequence. As concluded by Popielski, 
Dziembowski and Cassisi (2000) such a~wide spread of periods probably 
indicates wide spread of metal content. 

\vspace*{5mm}
\Section{Completeness of the Catalog}
The probability that a given RR~Lyr star has been included to our catalog 
depends on several factors like its amplitude, luminosity and crowding in its 
vicinity. The completeness of the catalog was estimated in similar way as the 
completeness of the catalog of RR~Lyr stars in the Small Magellanic Cloud 
(Paper I), \ie by comparison of objects detected in the overlapping regions 
between the neighboring fields. Twenty three such regions exist between our 
fields, allowing to perform 46 tests of paring objects from a given and 
adjacent fields. In total, 1079 stars from our lists should be paired with 
counterparts in the overlapping fields. We found counterparts in 1022 cases, 
which yields the completeness of our full sample equal to about 95\%. 

One should, however, remember that there is a number of variables which were 
missed in both overlapping fields. The regions near the edges of the fields 
are biased by smaller number of measurements, due to imperfections in the 
telescope pointing, what reduces the completeness of the catalog in the edge 
regions. We checked carefully all instances of unpaired objects. Counterparts 
of ten unpaired objects have the number of observations smaller than 100 and 
thus they were not searched for variability. More than 20 missed stars are 
located in the cores of dense star clusters or were blended with a bright 
star. That means that the completeness of the RR~Lyr sample detected in the 
star clusters is much lower than for the remaining regions. It is also 
possible that some RR~Lyr stars located close to the line-of-sight of bright 
stars could be missed. 

Additionally, we compared our sample to the MACHO list of 8654 RR~Lyr stars 
from the LMC, retrived from the MACHO archive
\vskip1pt
\centerline{({\it http://wwwmacho.mcmaster.ca/Data/MachoData.html}).}
\vskip1pt
3257 stars from the MACHO 
list could potentially be found in our fields. We did not find counterparts 
for 74 objects. Most of the missed stars were not RR~Lyr stars. The real 
RR~Lyr stars that were not identified in our sample were usually located very 
close to the edge of the frame and therefore they were biased by small number 
of observations. 

\Section{Summary}
In this paper we presented a huge sample of 7612 RR~Lyr stars found in the 
OGLE-II fields in the LMC. The mean parameters of all stars and individual 
measurements spanning four years of OGLE-II phase of the project are now 
available to the astronomical community. This is an ideal and unique dataset 
for further extensive studies of RR~Lyr stars. 

The third phase of the OGLE project (OGLE-III) started in June 2001 (Udalski 
\etal 2002) after significant increase of the observing capability. With the 
new mosaic CCD camera with the field of view increased to ${35\arcm\times 
35\zdot\arcm5}$ and sampling 0.25 arcsec/pixel almost entire Magellanic Clouds are 
now regularly monitored. Therefore, in the time scale of a few years complete 
catalogs of the vast majority of variable stars from both Magellanic Clouds 
should be released by the OGLE project. 

\vspace*{3mm}
\Acknow{We are grateful to Prof.\ B.\ Paczy\'nski for constant support and 
many useful suggestions during preparation of this paper. We thank Prof.\ W.\ 
Dziembowski for discussions and many important remarks. Also, we would like to 
thank Mr.\ Z.\ Ko{\l}aczkowski for help in estimation of the catalog 
completeness and providing the {\sc Fnpeaks} software. We are very grateful to 
Prof.\ G.\ Kov{\'a}cs for valuable comments to this work. The paper was partly 
supported by the Polish KBN grant 2P03D02124 to A.\ Udalski. Partial support 
to the OGLE project was provided with the NSF grant AST-0204908 and NASA grant 
NAG5-12212 to B.~Paczy\'nski. 

This paper utilizes public domain data originally obtained by the MACHO 
Project, whose work was performed under the joint auspices of the U.S. 
Department of Energy, National Security Administration by the University of 
California, Lawrence Livermore National Laboratory under contract No. 
W-7405-Eng-48, the National Science Foundation through the Center for Particle 
Astrophysics of the University of California under cooperative agreement 
AST-8809616, and the Mount Stromlo and Siding Spring Observatory, part of the 
Australian National University.}

\end{document}